\documentclass[preprint]{aastex}

\usepackage{graphicx}

\usepackage{natbib}

\newcommand{\be}{\begin{equation}}
\newcommand{\ee}{\end{equation}}
\newcommand{\nn}{\mbox{} \nonumber \\ \mbox{} }
\newcommand{\ba}{\begin{eqnarray}}
\newcommand{\ea}{\end{eqnarray}}

\newcommand{\Alfven}{Alfv\'{e}n\,}

\newcommand\eg{\textit{e.g.}}
\newcommand\cf{\textit{cf.}}

\newcommand{\Bf}{{magnetic field}}

\begin{document}

\title{Dynamics  of strongly  magnetized ejecta in Gamma Ray Bursts}

\author{Maxim Lyutikov\\
Department of Physics, Purdue University, \\
 525 Northwestern Avenue,
West Lafayette, IN
47907-2036 }

\begin{abstract}
We consider dynamical scales in magnetized GRB outflows, using the solutions to the Riemann problem of expanding arbitrarily magnetized outflows (Lyutikov 2010). For high ejecta magnetization, the   behavior of the forward shock closely resembles the so-called thick shell regime of  the hydrodynamical expansion.  The exception is at small radii, where the motion of the  forward shock is determined by the dynamics of {\it subsonic} relativistic outflows.  The  behaviors of the reverse shock is different in fluid and magnetized cases:  in the latter case, even for  medium magnetization,  $\sigma \sim 1$,  the reverse shock  forms at fairly large distances,  and may never form in a wind-type external density profile.
 \end{abstract}
\maketitle

\section{Introduction}

Magnetic fields  may play an important dynamical role in the GRB  outflows \cite[\eg][]{LyutikovJPh,Lyutikov:2009}. They may power the relativistic outflow through \cite{BlandfordZnajek} process \citep[\eg][]{Komissarov05}, and contribute to particle acceleration in the emission regions.
In this paper we discuss the dynamics of the relativistic, strongly magnetized   ejecta.
The results  are based on an exact solution of  a one-dimensional Riemann problem of expansion of a cold,  strongly magnetized into vacuum and into external medium of density $\rho_{\rm ex}$
  (Lyutikov, submitted); they are   reviewed in \S \ref{Riemann}.

In application to GRBs, we assumes that the central engine produces   jet  with density  $\rho_{0}$ and  magnetization $\sigma$ ($\sigma=B_0^2/\rho_{0}$; \Bf\ is normalized by $\sqrt{4 \pi} $), moving with Lorentz factor  $\gamma_w\gg 1$. In fact, parameters $\gamma_w$ and $\sigma$ are not always independent quantities: at small radii, when the motion of the ejecta is subsonic, they should be determined together with the motion of the boundary, see \S  \ref{subsonic}. 
In a supersonic regime, relation between $\gamma_w$ and $\sigma$ depends on the details of the flow acceleration (\eg\ in conical flows we expect 
$\gamma_w \sim \sqrt{\sigma}$). For generality, we do not assume any relationship between $\sigma$ and $\gamma_w$. 
The ejecta is moving into external density $\rho_{\rm ex}$. 

         \section{Riemann problem for relativistic expansion of magnetized gas}
\label{Riemann}

\subsection{Simple waves  and forward shock dynamics}

      Let us assume that the  jet  plasma is moving with velocity $\beta_w$ towards the external medium.        We   found 
  (Lyutikov, submitted) 
      exact self-similar  solution of relativistic Riemann problem for the  expansion  of  cold plasma with density $\rho_{0}$ and \Bf\ $B_0$ (magnetization parameter $\sigma=B_0^2/\rho_{0}$; \Bf\ is normalized by $\sqrt{4 \pi} $),   moving  initially with velocity $v_w$ towards the vacuum interface
 \ba &&
\delta_\beta = \delta_\eta^{2/3} \delta_{A,0}^{2/3} \delta_w ^{1/3}
\nn &&
\delta_A ={ \delta_{A,0}^{2/3}  \delta_w ^{1/3} \over \delta_\eta^{1/3}}
\label{main}
\ea
where  the Doppler factors $\delta _a= \sqrt{(1+\beta_a)/ ( 1-\beta_a)} $  are defined in terms of  the plasma velocity $\beta$, local \Alfven velocity $ \beta_{A}$, self-similar parameter $\eta= z/t$, initial wind velocity $\beta_w$ and the 
   \Alfven velocity in the undisturbed plasma $ \beta_{A,0}= \sqrt{ \sigma/(1+\sigma)}$.
These equations give the velocity $\beta$, density $\rho = U_A^2 \rho_0/\sigma$ ($U_A= \beta_{A} /\sqrt{1-\beta_{A} ^2}$) and proper  \Bf, $B =(\rho/\rho_0) B_0$ as a function of the self-similar variable $\eta= z/t$ (expansion of plasma  starts at $t=0, z=0$ and proceeds into positive direction $z>0$).  We stress that these solutions are exact, no assumptions about the value of the parameter $\sigma$ and velocity $v_w$ were made.

Particularly simple relations are obtained for plasma  initially at rest expanding into vacuum  $\beta_w=0, \,  \delta _\beta=1$ 
(Lyutikov, submitted). 
   The flow accelerates  from rest towards the vacuum interface.
The bulk of the flow is moving with Lorentz factor $\gamma' \sim \sigma^{1/3} $. The flow becomes supersonic  at $\eta =0$, at which point 
      $\gamma'= (\sigma/2)^{1/3}$.    The  vacuum interface moves  with Lorentz factor $\gamma_{vac} '= 1+2 \sigma$.
      In the observer frame the vacuum interface is moving with $\delta _\eta=  \delta_{A,0}^{2} \delta_w$, which 
       in the limit $\sigma, \, \gamma_w \gg1$ this gives
\be
\gamma_{vac} = 4 \gamma_w  \sigma
\label{gammavac}
\ee
     As the flow expands, the local magnetization 
     \be
     \sigma_{loc} = {  B^2 \over \rho}=  \left({ \delta_{A,0}^{2/3}  \over  \delta_\eta^{1/3}} - { \delta_\eta^{1/3} \over  \delta_{A,0}^{2/3} }\right) 
     \label{sigmaloc1}
     \ee
     decreases.     At the sonic point $   \sigma_{loc}  = (\sigma/2)^{2/3}$.

If there is an outside medium with density $ \rho_{\rm ex}$, we may identify two expansion regimes. For relativistically  strong forward shocks, so that the post-shock pressure is much larger than density,  the Lorentz factor of the CD is 
 \be
\gamma_{CD} =
\left({3 B_0^2 \gamma_w^2 \over 8 \rho_{\rm ex}} \right)^{1/4}  \approx \left({ L\over \rho_{\rm ex} c^3 } \right)^{1/4}  r^{-1/2}
\label{gammaCD}
\ee
(the last approximation assumes $\sigma \gg 1$). 
For weak forward shocks  the velocity of the CD approaches the expansion velocity into vacuum $\gamma_{vac}$, Eq. (\ref{gammavac}). The transition between the relativistically  strong  and weak shocks occurs  for 
\be
\sigma_{crit} = \left({ 3 \over 2048 \gamma_w^2} {\rho_0 \over  \rho_{\rm ex}} \right)^{1/3}
\label{rhoex1}
\ee
For $\sigma < \sigma_{crit}$, the forward shock is weak.
   
\subsection{Existence of the reverse shock}

For cold unmagnetized jets the reverse shock always exists; it is weak for $\gamma_w \leq \sqrt{\rho_0 / \rho_{\rm ex}}$ and strong otherwise \citep{Sari95}.
For magnetized jets the conditions for existence of a reverse shock are more complicated \citep[see also][]{2008A&A...478..747G,Mizuno}.
There are, in fact,  two somewhat different  regimes for the existence of a RS in highly  magnetized outflows. First,  if  ejecta is supersonic with respect to the CD (in term of  Riemann waves, this  transition corresponds to the case when the location of the FS   coincides with the location of the rarefaction wave), a strong RS  must forms. Secondly, if  the ejecta is subsonic with respect to the CD, but moves with velocity higher than the CD,  slowing of the ejecta is achieved by a compression wave, which may or may not turn into a reverse shock. One dimensional compression waves are always unstable to shock formation \citep{LLIV}. In contrast,  multidimensional subsonic outflow need not form shocks. 
So, formally,  the condition for reverse shock is
$\gamma_w > \gamma_{CD}$, but in the range $ \gamma_{CD} < \gamma_w < 2 \gamma_w \sqrt{\sigma}$ the RS may not form, if a more complicated flow patters are allowed. In any case, the RS shock, even if it exists, is weak in this regime.

Conditions for {\it strong} reverse shock (which implies a highly  supersonic flow, with velocity much larger than the \Alfven velocity in the upstream plasma)  were derived by \cite{kc84}.
In the frame of the CD, the reverse shock is moving with  \citep{kc84}
\be 
\beta_{RS} ^{\prime , 2}
   \approx 1-{1 \over  \sigma}, \,\mbox{ for} \sigma \gg 1
   \ee
If $\gamma_w \geq  \gamma_{CD}$, the reverse shock is weak (if it exists) and one should use a more detailed calculations of the dynamics of perpendicular shocks of arbitrary strength.  
 
Thus, for 
$\sigma \geq 1$, the existence of strong RS 
 requires $\gamma_w > 2 \gamma_{CD} \sqrt{\sigma}$, which using Eq. (\ref{gammaCD}) gives
\be
\gamma_w > \sqrt{6} \sqrt{ \rho_0 \over  \rho_{\rm ex}} \sigma^{3/2}, 
\label{RS1}
\ee
while a weak RS may exist for $ \gamma_{CD} < \gamma_w < 2 \gamma_w \sqrt{\sigma}$:
\be
\sqrt{3\over 8} \sqrt{ \rho_0 \over  \rho_{\rm ex}} \sqrt{\sigma} < \gamma_w <  \sqrt{6} \sqrt{ \rho_0 \over  \rho_{\rm ex}} \sigma^{3/2},
\label{RSweak}
\ee

The Lorentz factor of a strong  RS with respect to the contact discontinuity is $\sqrt{\sigma}$ (assuming $\sigma \gg 1$).   The
 Lorentz factor of the reverse shock in the frame  of stationary external medium is then
\be
\gamma_{RS}= {1\over 2} \left( { \sqrt{\sigma} \over \gamma_{CD} } + {\gamma_{CD}  \over \sqrt{\sigma} } \right)  =\left\{
\begin{array}{ll}
 \left( {3 \over 128} \right)^{1/4} { \sqrt{\gamma_w}   \over  \sigma^{1/4}} \left( {\rho_{0} \over \rho_{\rm ex}}\right)^{1/4} & \mbox{if $\gamma_{CD}   \gg  \sqrt{\sigma}$}\\
 { \sigma^{1/4} \over 6^{1/4} \sqrt{\gamma_w} }  \left( {\rho_{\rm ex} \over \rho_{0}}\right)^{1/4} & \mbox{if $\gamma_{CD}   \ll  \sqrt{\sigma}$}
\end{array}
\right.
\label{1}
\ee
 The two cases in Eq. (\ref{1}) correspond to RS moving in the same direction as the CD, $\gamma_{CD}  > \sqrt{\sigma}$, and the RS moving in the opposite direction than the CD, $\gamma_{CD} < \sqrt{\sigma}$. Condition $\gamma_{CD}  = \sqrt{\sigma}$ gives
 \be
 \gamma_w = \sqrt{8 \over 3}   \sqrt{\rho_{\rm ex} \over \rho_0} \sqrt{\sigma}
 \ee
 In this case the reverse shock is stationary  in the frame  of the external medium.


Let us summarize the main results. If the  the ratio of ejecta density to external density  is $f=  {\rho_{0} / \rho_{\rm ex}}$, then  weak RS can form for $\gamma_w \geq \gamma_{CD}$, which gives $\gamma_w \sim \sqrt{ f \sigma}$;  strong RS forms for $\gamma_w > \sqrt{\sigma} \gamma_{CD}$,
 $\gamma_w \geq  \sqrt{ f \sigma^3}$.
 RS shock propagates in the forward direction for $\gamma_{CD} > \sqrt{\sigma}$, $\gamma_w> \sqrt{\sigma /f}$. Forwards shock is relativistically weak for $\gamma_{CD} \geq \sigma$, $\gamma_w > \sigma^{3/2} /\sqrt{f}$  and becomes non-relativistic for $\gamma_{CD} \sim 1$, $\gamma_w < 1/\sqrt{\sigma f}$.

 \section{Dynamics of magnetized flows in GRBs}

In this section we apply the previous relationships to consider dynamics of  magnetized  flows in GRBs,  generalizing  discussion of  \cite{Sari95} to  strongly magnetized flows. We will derive main results in a thin shell approximation (not to be confused with a thin shell case, see below), assuming that the distances between the forward shock, the contact discontinuity and the reverse shock are small.  The velocity of the shocks and contact discontinuity are determined from the local force balance conditions. More precisely, they are determined by the {\it local} solutions to the Riemann problem of the decay of the discontinuity of the flow: there is no memory in the flow. 
Thin shell approximation is likely to be applicable, since for reasonable GRB parameters  the reverse shock never stalls while expansion is relativistic, see discussion after Eq. (\ref{RSstall}).



The ejecta flow  is taken to expands conically and carrying toroidal \Bf. 
We assume that the central source operates for time $\Delta t_s = \Delta/c$  ($ \Delta$ is the initial  width of the launched shell) and produces a wind with magnetization $\sigma \gg 1$ (magnetization $ \sigma= B^2/\rho$ is twice  the ratio of magnetic to particle energy in plasma frame; \Bf\ is normalized by $ \sqrt{4 \pi}$), moving with the Lorentz factor $\gamma_w$.   For spherical expansion (expansion along conical surfaces) of magnetized flows into vacuum, the magnetization parameter $\sigma$ remains constant outside the fast magnetosonic surface \citep{Michel73,Vlahakis03}. 
As we will see, the above assumption (that the central source produces a flow with a given $\gamma_w$ and $\sigma$) is not self-consistent for small radii, where the reverse shock does not form. In this case of subsonic expansion,  the flow dynamics cannot be specified {\it ad hoc}: it  needs to be determined self-consistently with the motion of the boundaries.

The wind luminosity is assumed to be  $L_{\rm iso} = E_{\rm iso}/ \Delta t_s$ where $E_{\rm iso}$ is the isotropic equivalent energy released by the central source.
Luminosity is produced in a form of Poynting and particle  fluxes 
\be 
L = 4 \pi r^2 \gamma_w^2 ( B_0^2 + \rho_0)  = 4 \pi r^2 \gamma_w^2 B_0^2 { 1+\sigma \over \sigma}
\ee
We are interested in the case $\sigma \geq 1$.
For numerical estimates we will use the typical values for long GRBs: $L_{\rm iso} =10^{51} $ erg s$^{-1}$, $\Delta t_s=100 $ s, $E_{\rm iso}=10^{53} $ erg, 
$\gamma_w = 300$. External density is $\rho_{\rm ex} = m_p n$.

\subsection{Forward shock dynamics}

In case of magnetized ejecta, as well as in the hydrodynamical case  \citep{Sari95}, the important scales in the problem (Sedov scale  $l_S$ (\ref{ls}),
energy scale $r_E$  (\ref{rdec}), reverse shock formation scale $r_N$ (\ref{rN}), reverse shock crossing scale $r_\Delta$ (\ref{RDelta}) and 
spreading distance  $r_S$ (\ref{RS}))  are related by a quantity \citep{Sari95}
\be
\xi=   \sqrt{ l_S \over  \Delta} \gamma_w^{-4/3} 
,\,\,\,
r_N/\xi = r_E= \sqrt{\xi} r_\Delta  = \xi^2 r_s
\label{xi}
\ee
In the hydrodynamical case, the parameter $\xi$ determines whether the reverse shock  and the rarefaction wave reach the whole ejecta  before most of the energy is transferred to the forward shock, $\xi > 1$, or later, $\xi < 1$. The dynamics of  magnetized ejecta generally follows  the hydrodynamic thick case, though the meaning of some radii change (\eg, in case of strongly magnetized ejecta $r_N$ is the scale of RS formation). 

There is a number of typical radii  where dynamics of the  outflow changes. 
There is Sedov radius 
\be
l_S \sim \left( { E_{\rm iso} \over  \rho_{\rm ex} c^2  } \right)^{1/3} =  4 \times 10^{18}\, {\rm cm}\,  n^{-1/3}
\label{ls}
\ee
where the ejecta and the swept-up ISM material become non-relativistic.

There is   radius $r_{E} $,  where the ejecta deposits approximately half of the energy or momentum to the external medium. For supersonic flows,   which reached terminal Lorentz factor $\gamma_w$, equating  energy in the shocked medium $\gamma_w^2 \rho_{\rm ex} c^2  r_{E} ^3$ to the total energy $E_{\rm iso}$, gives \citep{MeszarosRees92}
\be
r_{E} \sim \left( { E_{\rm iso} \over   \gamma_w^2 \rho_{\rm ex} c^2 } \right)^{1/3} = {l_S \over \gamma_w^{2/3}} = 9 \times 10^{16} \, {\rm cm}\,  n^{-1/3}
\label{rdec}
\ee 
 $r_{E} $   depends exclusively on the total energy of the explosion and not on its form (magnetic or baryonic).  
 For radii smaller  than $r_E$ the ejecta's  and the forward shocks' Lorentz factors remain constant and equal to the initial Lorentz factor $\gamma_w$. 
 For larger radii the flow enters the self-similar   Sedov-Blandford-McKee stage, with Lorentz factor  decreasing according to 
\be
\gamma = \left( {l_s \over r} \right)^{3/2}
\label{McKee}
\ee

In case of pure baryonic flow, and only in that case, $r_{E}$ is also the radius when the swept-up mass  equals the ejecta  mass divided by  $\gamma_w$
\be
r_{M} \sim \left( { M_0 \over  \gamma_w \rho_{\rm ex} c^2  } \right)^{1/3}
= \left( { E_K  \over \gamma_w^2 \rho_{\rm ex} c^2  } \right)^{1/3}
\ee
Here
 $E_K=E_{\rm iso}/(1+\sigma)$ is the energy associated with
bulk motion of matter.
Only in the case of zero magnetization   $r_{E}$  equals  $r_{M}$,  since
in that case $  E_{\rm iso}=E_K  = M_0 \gamma_w$. 
\footnote{The two radii  $r_{M}$ and $ r_{E}$ were confused by 
\cite{ZhangK}, who ''define the deceleration
radius using $E_K$ alone [] where the
fireball collects $1/\gamma_w$ of fireball rest mass''.  According to \cite{ZhangK}   "only the kinetic energy of the baryonic 
component ($E_K$) defines the afterglow level", while magnetic energy is transferred at unspecified "later" time. 
This is  incorrect \citep{LyutikovZhang}.}
For highly magnetized  outflow
$
{r_{M} \over r_{E}} = \sigma^{-1/3} \ll 1
$.

The above description of the forward shock dynamics is, in fact, applicable only in the so called  thin shell case,  $\xi >1$ (see Eq. (\ref{xi}). In this case the reverse shock quickly crosses the ejecta, which becomes causally connected so that all of the ejecta interacts with the external medium. 
Alternatively, 
 in the thick shell case,  $\xi<1$ (see Eq. (\ref{xi}),  the  reverse shock  does not have time to cross the ejecta before  the causally connected shocked part starts to decelerate at smaller radius $r_N$,  Eq. (\ref{rN}).  The Lorentz factor starts decreasing, but since new material and new momentum is being added to the shocked part of the ejecta, the ejecta and the forward shock behave effectively as a self-similar shock with energy supply
 \be
 \gamma = \left( {L  \over  \rho_{\rm ex}  c^2} \right)^{1/4} {1\over \sqrt{r}} = {l_s^{3/4} \over \Delta ^{1/4} \sqrt{r}}
\label{GammaISM}
\ee
Since in the thick shell case the Lorentz factor starts to decelerate earlier than in the thin shell case,  the rate of energy transfer to the external medium is smaller, so that the ejecta 
 gives most of its initial energy to the ISM at larger distances $r_\Delta > r_E$   \citep{Sari95}.
 
 \subsection{Formation and dynamics  of  the reverse shock}
  
The  weak reverse shock may form at (see Eq. (\ref{RSweak}))
\be
r _{N}=  {1  \over \gamma_w^2 }    \sqrt{3 L\over  2 \pi \rho_{ISM} c^{3}}  
 \approx {1  \over \gamma_w^2 }   {  l_S^{3/2}  \over  \sqrt{\Delta} } =
 10^{16} \, {\rm cm}   \, n^{-1/2}
 \label{rN}
\ee
RS becomes strong at $r _{RS, {\rm strong}} \sim \sigma r _{N}$ (see Eq. (\ref{RS1})).


 If the outside medium is stellar wind, strong  RS forms immediately if
 \be
 \gamma_w > \left({3 \over 2 \pi }  { L v_{wind} \sigma^2 \over c^3 \dot{M}} \right)^{1/4} = 220  \sigma^{1/2} L_{51}^{1/4}  v_{wind, 8}^{1/4} 
 \left( { \dot{M} \over 10^{-8} M_\odot/{\rm yr} } \right)^{-1/4}
  \label{rss1}
 \ee
 where $ v_{wind, 8}$ is the velocity of the progenitors wind in thousands kilometers per second.
 For smaller $\gamma_w$, no RS forms ever  (for weak shocks, one should put $\sigma \rightarrow 1$).

The   reverse shock could  stall (in the observer frame)  at (assuming $\sigma  \gg 1$)
\be
r_{RS,stall}={1\over \sigma}  \sqrt{ 3 L \over 8 \rho_{\rm ex} c^3} \sim {1\over \sigma} {l_S^{3/2} \over \sqrt{\Delta}} = 3  \times 10^{21} {\rm cm}  n^{-1/2} \sigma^{-1}
\label{RSstall}
\ee
Since  $r_{RS,stall}$ is typically larger than $l_S$, RS does not stall during the relativistic expansion phase; thus,  the thin shell approximation is generally applicable. 

In the unmagnetized case, an important quantity is the radius when
 the RS crosses the ejecta 
\be
r_\Delta \sim \left({ E \Delta \over \rho_{\rm ex} c^2} \right)^{1/4} = l_S ^{3/4} \Delta ^{1/4} =  10^{17} \, {\rm cm} \, n^{-1/4}
\label{RDelta}
\ee
In the magnetized case, $r_\Delta$ is still a good approximation for the  RS crossing radius, but with two cavities. First, a delayed onset of the RS, see (\ref{rN}),  delays the RS  crossing moment.  Since $r_N /r_\Delta = \xi^{3/2}$, this delay is not important for $\xi < 1$ (the thick shell case, generally applicable to the magnetized ejecta). Also, for subsonic outflows (see below), $r_\Delta$ is the distance where the back of the outflow catches with the CD, see \S \ref{subsonic}.

Second, magnetized shell is necessarily expanding, so that the tail part of the flow is moving with $\gamma \sim \gamma_{CD}/(2 \sqrt{\sigma})$. 
The typical shell spreading distance is 
\be
r_{S} \sim {\Delta   } \gamma_w^2
=3 \times 10^{17} {\rm cm}
\label{RS}
\ee 
Since the spreading occurs with Alfven velocity, the 
 tail part of the flow catches with the CD in the Blandford-McKee phase at $r_{tail} \sim \sqrt{\sigma} r_\Delta$. 
 We stress that spreading of magnetic shell, unlike of the cold baryonic shell, is unavoidable consequence of the high internal pressure (spreading of cold baryonic shell requires internal motion).
This is the reason why magnetic outflows are similar to the thick shell case of the baryonic outflows. 

\subsection{Dynamics of subsonic expansion}
\label{subsonic}
 
Thus, for   a given $\gamma_w$ and $\sigma$, at distances $r< r_N$ a rarefaction wave is launched into the flow, while the flow accelerates  to Lorentz factors larger than $\gamma_w$. This implies that our assumption that a flow has a given    $\gamma_w$ and $\sigma$ is not justified at $r< r_N$: at these distances the flow is effectively subsonic and its dynamics needs to be solved self-consistently, taking into account interaction with the external medium. 
The subsonic outflow may be considered as a collection of outgoing fast magnetosonic waves propagating from the central source, which constantly 
re-energize the FS.   The outflow  may be separated into two stages, which we   will call
''early'' and ''late'', depending on whether or not most of the
fast waves emitted by the central source have caught up with the
CD and their energy has been given 
to the circumburst medium. The transition between two stages occurs  at the moment, which
is similar to the shell crossing 
radius in the supersonic case, except that in the case of subsonic expansion the CD is decelerating 
all the time, but with different laws before and after the transition.

At the "early" stage 
 the  CD  is constantly re-energized by the
fast-magnetosonic waves propagating from the central source.
The   motion of the CD is determined by the
 luminosity 
at the retarded time $t'$:
\be
L_\Omega(t') \sim   \rho_{\rm ex} c^3 \gamma_{CD} ^4 r^2
\label{L}
\ee
(this is a condition of pressure balance between the wind and the ram pressure of ISM in the frame of the CD). 
For constant  luminosity  Eq. (\ref{L}) gives Eq. (\ref{GammaISM}) for the Lorentz factor of the CD.
This is exactly the same estimate as  for the intermediate scale in $\xi< 1$ supersonic flows, since $L_\Omega\sim E_{\rm iso} c/\Delta$; also
this is the same scaling as in the case of relativistic fluid reverse shock \citep{Sari97}. We stress that in the limit of strong FS,  Eq. (\ref{rhoex1}),  {\it the FS dynamics is  independent of the composition of the flow}, only the total power is important, Eq. (\ref{L}).

The early stage lasts 
 for  $r<r_\Delta$, Eq. (\ref{RDelta}).
At larger radii the flow enters the self-similar Sedov-Blandford-McKee stage.
At this stage, only a fraction of the shell  is interacting with the external medium, while 
the newly shocked ejecta material keeps adding energy and momentum to the shocked shell and the ISM, which evolve, effectively, as a flow with  energy supply.
Finally, for $r> r_\Delta$ the shock enters Blandford-McKee stage, with Lorentz factor given by Eq.  (\ref{McKee}).

 \begin{figure}[h!]
 \begin{center}
\includegraphics[width=.49\linewidth]{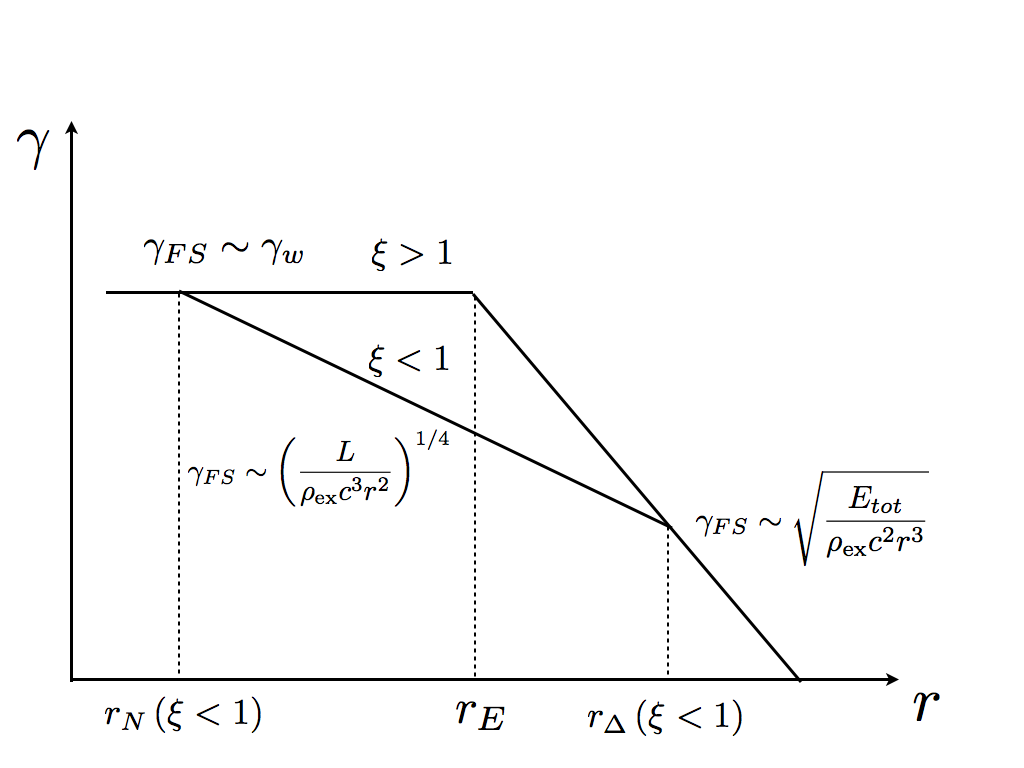}
\includegraphics[width=.49\linewidth]{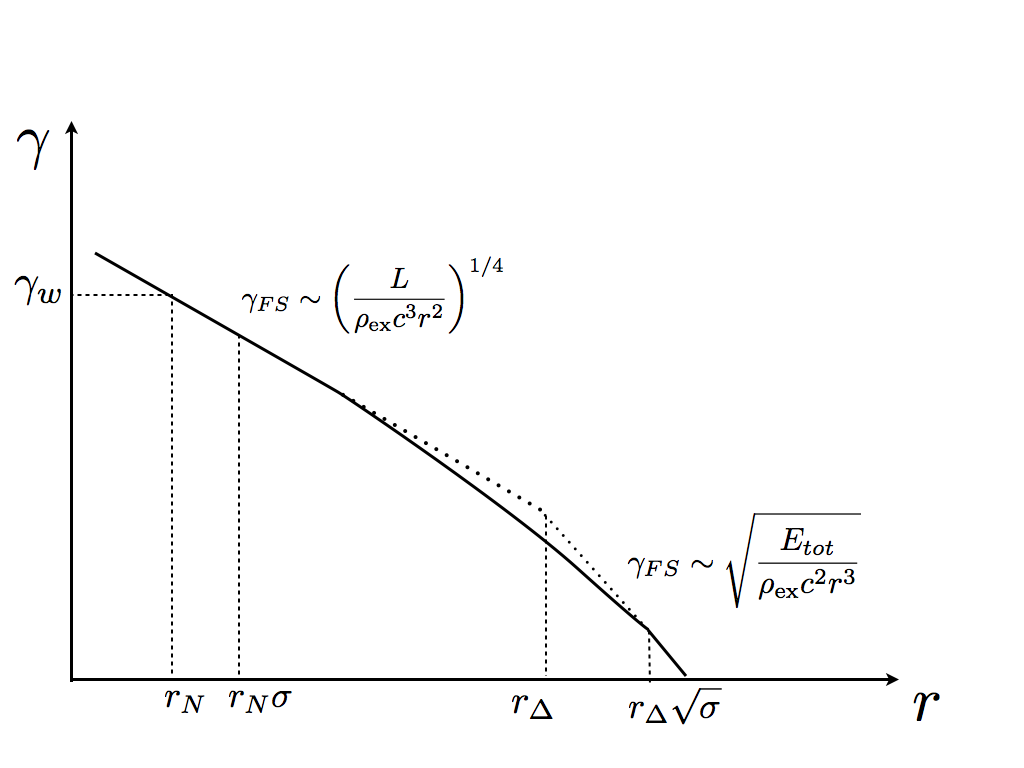}
\end{center}
\caption{Evolution of the Lorentz factor of the forward shock for matter-dominated  (Left panel) and Poynting flux-dominated models (Right panel). 
Matter-dominated  ejecta coasts with the injection Lorentz factor $\gamma_w$ until either $r_E$ (for $\xi > 1$, thin shell  case) or until $r_N$
(for $\xi < 1$, thick shell case). At $r_N$ reverse shock becomes strong.  At large radii ($r> r_E$ or $r> r_\Delta$)  the outflow enters the  Sedov-Blandford-McKee regime. 
For highly magnetized ejecta the  Lorentz factor of the CD and the FS 
initially decreases $\Gamma_{ISM}\propto r^{-1/2}$, changing to Sedov-Blandford-McKee regime $\Gamma_{ISM}\propto r^{-3/2}$  approximately at $r_\Delta$.  Reverse shock is launched at $r_N$  and becomes strong at $r_N \sigma$. Due to internal expansion of the magnetized shell, the back of the shell catches with the CD at distance $\sim r_\Delta \sqrt{\sigma}$. This is the reason why at distances close to $r_\Delta$ the Lorentz factor starts decreasing below the $r^{-1/2}$ law.}
\label{xiless1}
\end{figure}

\section{Discussion}

In this paper we discuss the dynamics of strongly magnetized   outflows in GRBs.  We find that the evolution of the forward shock driven by  strongly magnetized outflows are qualitatively the same as in the case of fluid shocks. The   definitions of radii  $r_N,\,r_E$ and $r_\Delta$   involve only the total energy of the ejecta, it's thickness and initial Lorentz factor, and {\it not} the information about it's content, \eg, parameter $\sigma$. The  typical radii (\ref{xi}) are the same for two flows \citep[\cf\ Eq.  (\ref{xi}) of the present paper and Eq. (9-10) of ][]{Sari95}.
These similarities may be understood, first, by noting that jump conditions in perpendicular magnetized shocks may be reduced to fluid shock jump conditions, with an appropriate choice of the equation of state, and, second, by the fact that the thin shell approximation is applicable in our case (so that the global  conservation of the toroidal magnetic flux, which modifies the global flow dynamics \citep{kc84},  is not important). Another   reason for this similarity is that \Bf\ behaves in many respects as a fluid with internal pressure. The only difference in the dynamics of the forward shocks driven by magnetized and fluid flows occurs for supersonic flows, $\gamma_w > \sqrt{\sigma} \gamma_{CD}$,  at very early stages  $r\leq r_N$ or $r\leq r_E$, see Fig. (\ref{xiless1}). Qualitatively, magnetized outflows are similar to thick shell hydrodynamic outflow, $\xi< 1$  at $r> r_N$.

Only at very early times, at $r< r_N$,  the forward shock bears information about anergy content: forward shock is coasting with $\gamma_w =$const in the fluid case and decelerating $\gamma \propto r^{-1/2}$ in the magnetized case. Dynamics of the reverse shock is quite different in case of high magnetization. First, the reverse shock forms at a finite distance from the source (Eq. \ref{rN}), and may not form at all in a wind environment,  (Eq. \ref{rss1}).  This fact may be related to observed paucity  of optical flashes in the  {\it Swift} era 
 \citep{Gomboc}. (The standard model had a clear prediction, of a bright optical flare with a definite decay properties \citep{1996ApJ...473..204S,1997ApJ...476..232M}. Though a flare closely resembling the predictions was indeed observed \citep[GRB990123,][]{1999MNRAS.306L..39M}, this was an exception.) 

In addition, at distances  $r_N< r< \sigma r_N$, where the RS is weak, the formation of  the RS shock depends on the details of the flow: RS forms if the flow is strictly radial, but need not to form if the the flow pattern is more complicated. We suggest that optical variability often seen in GRBs (\eg\ GRB021004 and most notoriously GRB080916C) is a reflection of the non-trivial flow patterns and the corresponding non-steady RS formation. 
Also, a recent detection of high polarization in optical \citep{2009Natur.462..767S} indicates a presence of an ordered \Bf\ in the ejecta.

I am greatly  thankful to  Dimitros Gianios, Sergey Komisarov and Alexandre Tchekhovskoy. 
\bibliographystyle{apj}
\bibliography{/Users/maxim/Home/Research/BibTex}

      \end{document}